\documentclass[authoryear,3p,times]{elsarticle}
\usepackage{natbib}
\usepackage{graphicx}
\usepackage{epsfig,subfigure}
\usepackage{amssymb}
\usepackage{amsthm}
\usepackage{amsmath}
\def\astrobj#1{#1}

\journal{New Astronomy}
\begin{document}
\begin{frontmatter}
\title{The photometric investigation of the newly discovered W UMa type binary system \astrobj{GSC 03122-02426}}

\author[1,2,3]{Zhou, X.}
\author[1,2,3]{Qian, S.-B.}
\author[1,2]{He, J.-J.}
\author[1,2]{Zhang, J.}
\author[1,2,3]{Zhang, B.}

\address[1]{Yunnan Observatories, Chinese Academy of Sciences, PO Box 110, 650216 Kunming, China}
\address[2]{Key Laboratory of the Structure and Evolution of Celestial Objects, Chinese Academy of Sciences, PO Box 110, 650216 Kunming, China}
\address[3]{Graduate University of the Chinese Academy of Sciences, Yuquan Road 19, Sijingshang Block, 100049 Beijing, China}
\cortext[cer]{Corresponding author. Tel.: +8613658806295 \\
E-mail address: zhouxiaophy@ynao.ac.cn}


\begin{abstract}
The $B$ $V$ $R_c$ $I_c$ bands light curves of the newly discovered binary system \astrobj{GSC 03122-02426} are obtained and analyzed using the Wilson-Devinney (W-D) code. The solutions suggest that the mass ratio of the binary system is $q = 2.70$ and the less massive component is $422K$ hotter than the more massive one. We conclude that \astrobj{GSC 03122-02426} is a W-subtype shallow contact (with a contact degree of $f = 15.3\,\%$) binary system. It may be a newly formed contact binary system which is just under geometrical contact and will evolve to be a thermal contact binary system. The high orbital inclination ($i = 81.6^{\circ}$) implies that \astrobj{GSC 03122-02426} is a total eclipsing binary system and the photometric parameters obtained by us are quite reliable. We also estimate the absolute physical parameters of the two components in \astrobj{GSC 03122-02426}, which will provide fundamental information for the research of contact binary systems. The formation and evolutionary scenario of \astrobj{GSC 03122-02426} is discussed.
\end{abstract}


\end{frontmatter}

\section{Introduction}
W UMa type binary systems are cool short-period binary systems in which both of its two components fill the critical Roche lobes and share a common convective envelope. Their EW-type light curves vary continuously and the depths of the primary and secondary minima are almost equal. It is found that about $95\,\%$ of the eclipsing binaries around the solar system are W UMa type contact binary systems \citep{1948HarMo...7..249S}. \citet{1970VA.....12..217B} divided the W UMa type binary systems into two subclasses: A-subtype and W-subtype. The A-subtype W UMa systems are also called early-type contact binaries with their spectral types usually earlier than F0 and the more massive stars are the hotter ones. The W-subtype systems are also called late type contact binaries with their spectral types to be F and G type or even later. The less massive components are the hotter ones in the W-subtype systems. The formation and evolutionary theories of contact binaries are still open issues. W UMa type systems are very important targets in this research field. Therefore, more and more fundamental information about W UMa type systems, both photometric and spectroscopic analysis, are urgently needed.

\astrobj{GSC 03122-02426}, also named \astrobj{UCAC4 652-064817} or \astrobj{2MASS J18454487+4017197}, is a newly discovered W UMa type contact binary system detected in the field of view of V563 Lyr. There is neither photometric solution nor spectroscopic observation about this target published before. In the present work, $B$ $V$ $R_c$ and $I_c$ bands light curves of \astrobj{GSC 03122-02426} are analyzed with the Wilson-Devinney (W-D) code for the first time. The basic physical parameters of this system is obtained and its formation and evolutionary scenario are figured out, which will improve our understanding on the W UMa type binary systems.

\section{The CCD photometric observations of \astrobj{GSC 03122-02426}}

The CCD photometric observations of the eclipsing binary \astrobj{GSC 03122-02426} were carried out over 7 nights on September and October 2012 (HJD 2456199 to HJD 2456207) using an SBIG ST-10XME CCD camera attached to the 0.4m Cassegrain reflecting telescope in the University of Athens Observatory, Greece \citep{2015IBVS.6199....3G}. During the observations, the broadband Johnson-Cousins $B$ $V$ $R_c$ $I_c$ filters were used. \astrobj{GSC 03122-00809} was chosen as the comparison star. A total of 358 points in $B$ filter, 395 in $V$ filter, 383 in $R_C$ filter and 417 in $I_C$ filter were collected, respectively. The coordinates and brightness (in $V$ band) of the variable, the comparison stars are listed in Table \ref{Coordinates}. The identification map of \astrobj{GSC 03122-02426} in the field of view of \astrobj{V563 Lyr} is shown in Fig. 1.

\begin{table}
\begin{center}
\caption{ Coordinates of \astrobj{GSC 03122-02426} and the comparison star}\label{Coordinates}
\begin{tabular}{ccccc}\hline\hline
Targets          &            name               & $\alpha_{2000}$        &  $\delta_{2000}$         &  $V_{mag}$      \\ \hline
Variable         &   \astrobj{GSC 03122-02426}   &$18^{h}45^{m}44^{s}.9$  & $+40^\circ17'20''.2$     &  $14.5$         \\
The comparison   &   \astrobj{GSC 03122-00809}   &$18^{h}45^{m}36^{s}.4$  & $+40^\circ08'05''.5$     &  $11.6$          \\
\hline\hline
\end{tabular}
\end{center}
\end{table}

\begin{figure}[!h]
\begin{center}
\includegraphics[width=10cm]{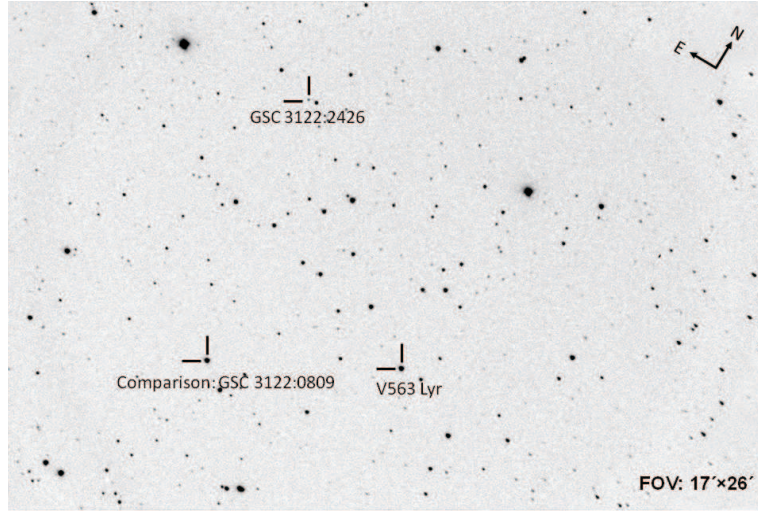}
\caption{ The identification map of \astrobj{GSC 03122-02426} in the field of view of \astrobj{V563 Lyr}}
\end{center}
\end{figure}

The first linear ephemeris of \astrobj{GSC 03122-02426} is obtained,
\begin{equation}
Min. I(HJD) = 2456200.2458(7) + 0^{d}.30050(9)\times{E}.
\end{equation}

\section{Photometric solutions of \astrobj{GSC 03122-02426}}
\astrobj{GSC 03122-02426} is a newly detected binary system with EW type light curves.  To understand its geometrical structure and evolutionary state, the $B$ $V$ $R_c$ and $I_c$ bands light curves covering complete phase ranges are analyzed simultaneously using the W-D code of version 2013 \citep{Wilson1971,Van2007,Wilson2010}. The phases applied in the W-D code are calculated with Equation (1).

Since \astrobj{GSC 03122-02426} is a newly discovered binary system, there isn't any information about spectral type on this target published before. Therefore, we can't determine the effective temperature of the binary system directly. Considering the color index of $J - H =0.42$ given by the fourth US Naval Observatory CCD Astrograph Catalog (UCAC4), we fix the effective temperature of the primary star (star eclipsed at the primary minimum light) to be $T_1 = 5230K$ \citep{Cox2000}. It means that \astrobj{GSC 03122-02426} is a late- type contact binary with convective envelopes. Thus the gravity-darkening coefficients $g_1=g_2=0.32$ \citep{1967ZA.....65...89L} and the bolometric albedo $A_1=A_2=0.5$ \citep{1969AcA....19..245R} are used. To account for the limb darkening in detail, logarithmic functions are used. The corresponding bolometric and passband-specific limb-darkening coefficients are chosen from \citet{1993AJ....106.2096V}'s table. Synchronous rotation and a circular orbit are assumed during the solution processes.

As there isn't any radial velocity study about \astrobj{GSC 03122-02426}, a $q$-search method is used to determine the initial input mass ratio for the light curves. During the calculating, it is found that the solution converges at mode 3 (contact model), and the adjustable parameters are: the mass ratio $q$ $(M_2/M_1)$; the orbital inclination $i$; the effective temperature of star 2 ($T_{2}$); the monochromatic luminosity of star 1 ( $L_{1B}$, $L_{1V}$, $L_{1R}$ and $L_{1I}$); the dimensionless potential of star 1 ($\Omega_{1}=\Omega_{2}$ in mode 3 for contact configuration). Solutions with mass ratio from 0.1 to 4.8 are investigated, and the relation between the resulting sum of weighted square deviations $\Sigma$ and $q$ is plotted in Fig. 2. The minimum values are found at $q$ = 2.65, which indicates that \astrobj{GSC 03122-02426} is a W-subtype \citep{1970VA.....12..217B} contact binary. Then $q$ = 2.65 is set as the initial value and considered as an adjustable parameter. The final photometric solutions are listed in Table 2 and the corresponding theoretical light curves are displayed in Fig. 3. The contact configurations of \astrobj{GSC 03122-02426} are displayed in Fig. 4.

\begin{table}[!h]
\caption{ Photometric solutions of \astrobj{GSC 03122-02426}}\label{phsolutions}
\begin{center}
\begin{tabular}{lllllllll}
\hline
Parameters                        & Values                         \\\hline
$T_{1}(K)   $                     & 5230(fixed)                   \\
$g_{1}$                           & 0.32(fixed)                   \\
$g_{2}$                           & 0.32(fixed)                   \\
$A_{1}$                           & 0.50(fixed)                   \\
$A_{2}$                           & 0.50(fixed)                   \\
q ($M_2/M_1$ )                    & 2.70($\pm0.07$)             \\
$i(^{\circ})$                     & 81.6($\pm0.5$)              \\
$\Omega_{1}=\Omega_{2}$           & 6.1168($\pm0.0883$)       \\
$T_{2}(K)$                        & 4808($\pm11$)                \\
$\Delta T(K)$                     & 422($\pm11$)  \\
$T_{2}/T_{1}$                     & 0.919($\pm0.002$)  \\
$L_{1}/(L_{1}+L_{2}$) (B)         & 0.432($\pm0.003$)           \\
$L_{1}/(L_{1}+L_{2}$) (V)         & 0.402($\pm0.003$)           \\
$L_{1}/(L_{1}+L_{2}$) (R)         & 0.380($\pm0.003$)           \\
$L_{1}/(L_{1}+L_{2}$) (I)         & 0.365($\pm0.003$)          \\
$r_{1}(pole)$                     & 0.2834($\pm0.0021$)           \\
$r_{1}(side)$                     & 0.2964($\pm0.0023$)          \\
$r_{1}(back)$                     & 0.3342($\pm0.0029$)           \\
$r_{2}(pole)$                     & 0.4469($\pm0.0080$)            \\
$r_{2}(side)$                     & 0.4798($\pm0.0110$)           \\
$r_{2}(back)$                     & 0.5090($\pm0.0151$)           \\
$f$                               & $15.3\,\%$($\pm$14.4\,\%$$)    \\
$\Sigma{\omega(O-C)^2}$           & 0.00826                     \\
\hline
\end{tabular}
\end{center}
\end{table}

\begin{figure}[!h]
\begin{center}
\includegraphics[width=12cm]{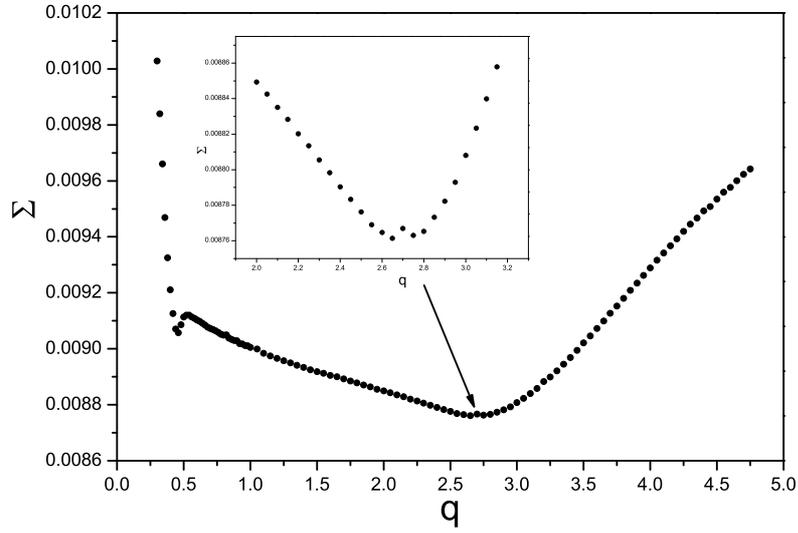}
\caption{ Relation between $\Sigma$ and $q$ for light curves of \astrobj{GSC 03122-02426}. $\Sigma$ is the resulting sum of weighted square deviations. It is shown that the minimum is obtained at q = 2.65.}
\end{center}
\end{figure}

\begin{figure}[!h]
\begin{center}
\includegraphics[width=10cm]{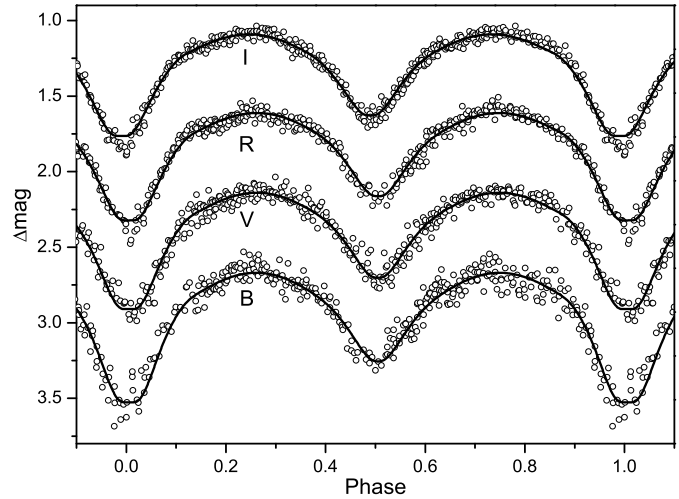}
\caption{Observed (open circles) and theoretical (solid lines) light curves in the $B$ $V$ $R_c$ and $I_c$ bands of \astrobj{GSC 03122-02426}. The standard deviation of the fitting residuals is 0.069 mag for $B$ band, 0.052 mag for $V$ band, 0.040 mag for $R_c$ band and 0.036 mag for $I_c$ band, respectively.}
\end{center}
\end{figure}

\begin{figure}[!h]
\begin{center}
\includegraphics[width=12cm]{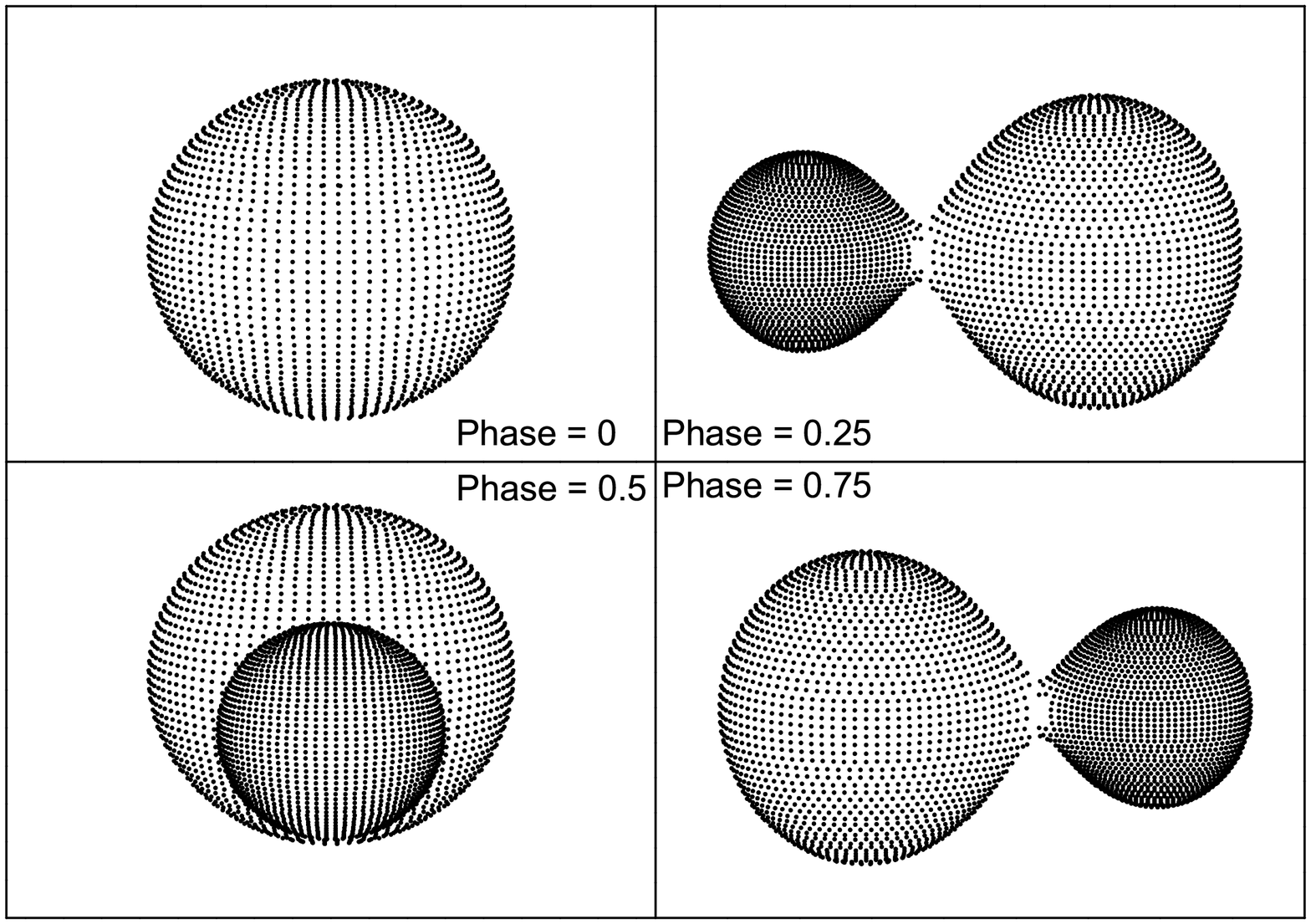}
\caption{Contact configurations of \astrobj{GSC 03122-02426} at phases 0.0, 0.25, 0.5 and 0.75.}
\end{center}
\end{figure}

\section{Discussions and Conclusions}
The CCD photometric light curves of \astrobj{GSC 03122-02426} in $B$ $V$ $R_c$ $I_c$ bands are analyzed simultaneously with the W-D code (Version 2013). Our solutions suggest that \astrobj{GSC 03122-02426} is a W-subtype shallow contact ($f = 15.3\,\%$) binary system with its mass ratio to be $q = 2.70$, where ``shallow" is defined when the degree of contact is lower than $20\,\%$ \citep{2016NewA...43....1L}. The temperature difference between its two components is $422K$ which may imply that \astrobj{GSC 03122-02426} is a newly formed contact binary system. It is just under geometrical contact and will evolve to be a thermal contact binary system. The less massive star contributes about $40\,\%$ of the total luminosity in the binary system. During the calculations, we also set $l_3$ as an adjustable parameter in the W-D code. However, we haven't detected the contribution of the third light in the light curves. The high orbital inclination ($i = 81.6^{\circ}$) implies that \astrobj{GSC 03122-02426} is a total eclipsing binary system and the photometric parameters obtained by us are quite reliable. By assuming that the primary star in \astrobj{GSC 03122-02426} is a normal main sequence star, we estimate the mass of the primary star to be $M_1 = 0.81M_\odot$ \citep{Cox2000}. Considering the mass ratio of $q = 2.70$ derived by us, the mass of the secondary star is calculated to be $M_2 = 2.19(\pm0.06)M_\odot$. The radii and luminosity of the two components are determined to be: $R_1 = 0.83(\pm0.01)R_\odot$, $R_2 = 1.30(\pm0.03)R_\odot$, $L_1 = 0.47(\pm0.01)L_\odot$ and $L_2 = 0.72(\pm0.02)L_\odot$. The orbital semi-major axis of \astrobj{GSC 03122-02426} is calculated to be $a = 2.72(\pm0.02)R_\odot$.

The formation and evolutionary theories of contact binaries are still unclear. The most possible scenario is that they are formed from initially detached binary systems through the angular momentum loss (AML) via the magnetic wind \citep{2016ApJ...817..133Z}. A late type star component may have played an important role in the formation process \citep{2013MNRAS.430.2029Y,2014MNRAS.437..185Y}. Therefore, the W UMa type systems are ideal targets to test this theory and the photometric analyses of the W UMa type systems will provide invaluable information on the research of star formation and evolution. As for \astrobj{GSC 03122-02426}, the temperature difference between its two components suggest that the system is under poor thermal contact. It may be a newly formed contact binary system. \citet{2010AJ....140..215Z} pointed out that the majority of W-subtype contact binaries are shallow contact systems. Our results for \astrobj{GSC 03122-02426} confirm the conclusion and imply that \astrobj{GSC 03122-02426} may be undergoing the thermal relaxation oscillation (TRO; \citet{1976ApJ...205..208L,1976ApJ...205..217F,1977MNRAS.179..359R}). The orbit of the binary system will shrink and expand under the control of the angular momentum loss (AML) and the thermal relaxation oscillation (TRO). The degree of contact will increase and the system will evolve into a fully thermal contact configuration as the increasing times of oscillating circles \citep{2001MNRAS.328..635Q,2001MNRAS.328..914Q}.

\section{Acknowledgments}
We thank K. Gazeas in Department of Astrophysics, Astronomy and Mechanics, National and Kapodistrian University of Athens for the data observed in University of Athens Observatory, Greece. This work is supported by the Chinese Natural Science Foundation (Grant No. 11133007, 11325315 and 11203066), the Strategic Priority Research Program ``The Emergence of Cosmological Structure'' of the Chinese Academy of Sciences (Grant No. XDB09010202) and the Science Foundation of Yunnan Province (Grant No. 2012HC011). This research has made use of the fourth US Naval Observatory CCD Astrograph Catalog (UCAC4).

\end{document}